\documentstyle[12pt]{article}

\newcommand{\be}{\begin{equation}}
\newcommand{\ee}{\end{equation}}
\newcommand{\bea}{\begin{eqnarray}}
\newcommand{\eea}{\end{eqnarray}}

\begin{document}

\large

\title{Quantum Mechanical Derivation of the Wallis Formula for $\pi$}

\begin{center}

\large {\bf 
Quantum Mechanical Derivation of the Wallis Formula for $\pi$}

\vskip .3cm

\normalsize

Tamar Friedmann\footnote{Email address: tfriedma@ur.rochester.edu} 

\small
{\it Department of Mathematics and Department of Physics and Astronomy \\ University of Rochester,  Rochester, N.Y. 14627}

\vskip .3cm

\normalsize
C. R. Hagen\footnote{Email address: hagen@pas.rochester.edu}
\small

{\it Department of Physics and Astronomy  \\
University of Rochester,  Rochester, N.Y. 14627 }
\end{center}

\begin{abstract}
A famous pre-Newtonian  formula for $\pi$ is obtained directly from the variational approach to the spectrum of the hydrogen atom in spaces of arbitrary dimensions greater than one, including the physical three dimensions. 
\end{abstract}
\vskip .7cm


\normalsize

The formula for $\pi$ as the infinite product 
\be \label{Wallis}{\pi \over 2}={2 \cdot 2\over 1\cdot 3}\; {4 \cdot 4\over 3\cdot 5}\; {6 \cdot 6\over 5\cdot 7}\cdots \ee 
was derived by John Wallis in 1655 \cite{Wal} (see also \cite{BBB}) by a method of successive interpolations. While several mathematical proofs of this formula have been put forth in the past
(many just in the last decade) using  probability \cite{M}, combinatorics and probability \cite{K}, geometric means \cite{Wast}, trigonometry \cite{SW, Weis}, and trigonometric integrals \cite{St}, 
there has not been in the literature a derivation of Eq. (\ref{Wallis}) that originates in physics, specifically in quantum mechanics. 

It is the purpose of this paper to show that 
this formula can in fact be derived from a variational computation of the spectrum of the hydrogen atom. The existence of such a derivation indicates that there are striking connections between well-established physics and pure mathematics \cite{land} that are remarkably beautiful yet still to be discovered.

The Schr{\"o}dinger equation for the hydrogen atom  is given by
$$ H \psi =\left ( -{ \hbar^2\over 2m}\nabla ^2 - {e^2\over r} \right ) \psi =E\psi ,
$$
with the corresponding radial equation obtained by separation of variables being
$$H(r) R(r)=\left [-{ \hbar^2\over 2m}\left ( {d^2\over dr^2}+{2\over r}{d\over dr}-{\ell(\ell +1)\over r^2}\right ) -{e^2\over r} \right ] R(r)=ER(r).
$$
Using the trial wave function
\be \label{trial} \psi_{\alpha \ell m} =r^\ell e^{-\alpha r^2} Y^m_\ell(\theta, \phi) , \ee
where $\alpha >0$ is a real parameter and the $Y^m_\ell (\theta, \phi)$ are the usual spherical harmonics,  the expectation value of the Hamiltonian is found to be given by
\bea \langle H\rangle _{\alpha  \ell}&\equiv& {\langle \psi_{\alpha \ell m}| H(r)  |\psi_{\alpha \ell m} \rangle \over \langle \psi_{\alpha \ell m} |\psi_{\alpha \ell m}\rangle} \nonumber \\
&=& \nonumber {\hbar ^2\over 2m}\left ( \ell +{3\over 2}\right ) 2\alpha -e^2{\Gamma(\ell +1)\over \Gamma (\ell+{3\over 2})}\sqrt{2\alpha}\; .
\eea
As a consequence of the $Y_\ell ^m(\theta, \phi)$ in Eq. (\ref{trial}), the function $\psi_{\alpha \ell m} $ is orthogonal to any energy and angular momentum eigenstate that has a value for angular momentum different from $\ell$. Therefore the minimization of  $\langle H\rangle _{\alpha \ell} $ with respect to $\alpha$, namely
\be \label{min}\langle H \rangle _{min}^{\ell} = -{me^4\over 2\hbar ^2}{1\over (\ell +{3\over 2})}\left [ {\Gamma(\ell +1)\over \Gamma (\ell +{3\over 2})}\right ]^2,\ee
gives an upper bound for the lowest energy state with the given value of $\ell$. 

The well-known exact result for the energy levels of hydrogen  is 
$$E_{n_r, \ell}=-{me^4\over 2\hbar ^2}{1\over (n_r+\ell+1)^2} \; ,
$$
where $n_r=0,1,2,\ldots$, so that the lowest energy eigenstate  for a given $\ell$ is the one with $n_r=0$, namely
$$E_{0,\ell}=-{me^4\over 2\hbar ^2}{1\over (\ell+1)^2} \; .$$
The accuracy of approximation  (\ref{min}) is thus displayed in the ratio
 $${\langle H \rangle _{min}^{\ell}\over   E_{0, \ell}}=
   {(\ell +1)^2\over (\ell +{3\over 2})}  \left [ {\Gamma(\ell +1) \over \Gamma(\ell +{3\over 2})}\right ] ^2,$$
 a quantity that approaches  unity with increasing $\ell$ 
\cite{elllimit}.  
This follows from the fact that in the large $\ell$ limit the trial solution and the exact result  correspond to strictly circular orbits. The circularity of the trial solution orbits at large $\ell$ is a consequence of the fact that the uncertainty in $r^2$, measured in units of mean square radius, is given by
$${[\langle r^4 \rangle _{\alpha\ell} -(\langle r^2 \rangle_{\alpha \ell})^2]^{1\over 2}\over \langle r^2 \rangle_{\alpha \ell}}= \left (\ell +{3\over 2}\right )^{-{1\over 2}},$$ 
which approaches 0 at large $\ell$. Both the trial solution orbits and the exact orbits are then identical to those of the Bohr model in the large $\ell$ limit, as expected from Bohr's correspondence principle.

Therefore one obtains the limit
\be \label{3d}\lim_{\ell \rightarrow \infty}
 { \langle H \rangle _{min}^{\ell}\over   E_{0, \ell}}=
   \lim_{\ell \rightarrow \infty} {(\ell +1)^2\over (\ell +{3\over 2})}  \left [ {\Gamma(\ell +1) \over \Gamma(\ell +{3\over 2})}\right ] ^2=1
\, .\ee
This can be seen to lead to the Wallis formula for $\pi$. To this end, one invokes  the relations $z\Gamma(z)=\Gamma(z+1)$, $\Gamma(\ell +1)=\ell!$, and $\Gamma ( {1\over 2})=\sqrt{\pi}$, which bring Eq. (\ref{3d}) to the form
$$  \lim _{\ell \rightarrow \infty} \left [ (\ell +1)!  \over \sqrt{\pi} \cdot {1\over 2}{3\over 2}{5 \over 2} \cdots {2\ell +1 \over 2} \right ] ^2 {1\over \ell +{3\over 2}}=1 
$$
or alternatively
\be \label{gotWall} {\pi \over 2}=\lim _{\ell \rightarrow \infty} \prod_{j=1}^{\ell +1} {(2j)(2j)\over (2j-1)(2j+1)} \; ,
\ee
i.e., the Wallis formula for $\pi$, as given by Eq. (\ref{Wallis}). 

The analogous computation in arbitrary dimensions also leads to the same formula, with slightly different forms for even and odd dimensions. 
The radial equation for the hydrogen atom in $N$ dimensions is \cite{Nieto}
\[ H_NR=\left [ -{\hbar ^2\over 2m}\left ( {d^2\over dr^2}+{N-1\over r}{d\over dr}-{\ell(\ell +N-2)\over r^2}\right ) -{e^2\over r} \right ] R(r)=ER(r),
\]
where $\hbar ^2 \ell (\ell+N-2)$, $\ell = 0,1,2, \ldots$  is the spectrum of the square of the angular momentum operator in $N$ dimensions \cite{Louck, FH}. 
The same trial wave function as in three dimensions with the $Y^m_\ell (\theta, \phi)$ of Eq. (\ref{trial}) replaced by its $N$-dimensional analog \cite{Nieto, Louck} gives
\be \langle H \rangle ^{N, \ell}_{min}=  -{me^4\over 2\hbar ^2}{1\over (\ell +{N\over 2})}\left [ {\Gamma(\ell +{N-1\over 2})\over \Gamma (\ell +{N\over 2})}\right ]^2.\ee
The exact result \cite{Nieto} 
 $$E^N_{n_r,\ell}=-{me^4\over 2\hbar ^2}{1\over (n_r+\ell+{N-1\over 2})^2}\; ,
$$
in the limit $\ell \rightarrow \infty$ with $n_r=0$ yields 
\be \label{Nd}\lim_{\ell \rightarrow \infty}
 { \langle H \rangle _{min}^{N,\ell}\over   E^N_{0, \ell}}=
   \lim_{\ell \rightarrow \infty} {(\ell +{N-1\over 2})^2\over (\ell +{N\over 2})}  \left [ {\Gamma(\ell +{N-1\over 2}) \over \Gamma(\ell +{N\over 2})}\right ] ^2=1.\ee

For $N=2k+1$, where $k$ is a positive integer (i.e., the case of odd dimensions),  this becomes
$$
   \lim_{\ell \rightarrow \infty} {(\ell +k)^2\over (\ell +k+{1\over 2})}  \left [ {\Gamma(\ell +k) \over \Gamma(\ell +k+ {1\over 2})}\right ] ^2=1$$
which is identical to Eq. (\ref{3d}) once the substitution $\ell \rightarrow \ell +k-1$ is made there, leading again to  the Wallis formula. For $N=2k$, where $k$ is again a positive integer (i.e., the case of even dimensions), Eq. (\ref{Nd}) becomes
$$
   \lim_{\ell \rightarrow \infty} {(\ell +k-{1\over 2})^2\over (\ell +k)}  \left [ {\Gamma(\ell +k-{1\over 2}) \over \Gamma(\ell +k)}\right ] ^2=1,$$
which is readily brought to the form
\be {2 \over \pi}=\lim _{\ell \rightarrow \infty} \prod_{j=1}^{\ell +1} { (2j-1)(2j+1) \over (2j)(2j)} ,
\ee
the reciprocal form of the Wallis formula. 


\renewcommand{\baselinestretch}{1}


\normalsize

\begin{thebibliography}{99}
\bibitem{Wal} J. Wallis, ``Arithmetica Infinitorum," Oxford, 1655.
\bibitem{BBB} L. Berggren, J. Borwein,  and P. Borwein, ``Pi: A source book," Springer-Verlag (1997). 
\bibitem{M} S. J. Miller, 
``A probabilistic proof of Wallis's formula for $\pi$,"
Am. Math. Monthly {\bf 115}, 740-745 (2008)  [arXiv:0709.2181].
\bibitem{K} M. Kovalyov,  
``Elementary combinatorial-probabilistic proof of the Wallis and Stirling formulas,"
J. Math. \&  Stat. {\bf 5}, 408-410 (2009).
\bibitem{Wast} J. W\"{a}stlund, 
``An elementary proof of Wallis's product formula for pi",
Link\"{o}ping studies in Mathematics, 2:1-5 (2005). 
\bibitem{SW} J. Sondow,  and E. W. Weisstein,  ``Wallis' formula." From {\it MathWorld} -- A Wolfram web resource. http://mathworld.wolfram.com/WallisFormula.html
\bibitem{Weis} E. W. Weisstein,  ``Pi formulas." From {\it MathWorld} -- A Wolfram web resource.  http://mathworld.wolfram.com/PiFormulas.html
\bibitem{St} J. Stewart, ``Calculus: Early Transcendentals," Cengage Learning, Seventh Edition 2012.

\bibitem{land}  For an example of such a connection, 
see P. T. Landsberg, 
``A thermodynamic proof of the inequality between arithmetic and geometric mean,"
Phys. Lett. A {\bf 67}, 1 (1978). 
\bibitem{elllimit} For example, the ratio is $.849, .906, .932,  .978, .998$ for $\ell = 0,1,2,  10, 100$.
\bibitem{Nieto} M. M. Nieto, 
``Hydrogen atom and relativistic pi-mesic atom in $N$-space dimensions,"
Am. J. Phys. {\bf 47}, 1067 (1979). 
\bibitem{Louck} J. D. Louck, \emph{Theory of Angular Momentum in N-Dimensional Space}, 
Los Alamos Scientific Laboratory monograph LA-2451 (LASL, Los Alamos, (1960));
J. D. Louck, J. Mol. Spec {\bf 4}, 298 (1960); 
J. D. Louck and H. W. Galbraith, Rev. Mod. Phys {\bf 44}, 540 (1972). 

\bibitem{FH} T. Friedmann and C. R.  Hagen,
 ``Group-theoretical derivation of angular momentum eigenvalues in spaces of arbitrary dimensions,”
J. Math. Phys. {\bf 53}, 122102 (2012) [arXiv:1211.1934]. 


\end{thebibliography}
\end{document}